\documentclass[preprint,aps,12pt,showpacs,nofootinbib,tightenlines,superscriptaddress]{revtex4}

\usepackage{amsmath}
\usepackage{graphicx}
\usepackage{amssymb}
\begin{document}
\def\intdk{\int\frac{d^4k}{(2\pi)^4}}
\def\sla{\hspace{-0.17cm}\slash}
\hfill

\title{Note on Higgs Decay into Two Photons $H\to \gamma\gamma$}
\author{Da Huang}\email{dahuang@itp.ac.cn}
\affiliation{Kavli Institute for Theoretical Physics China (KITPC)
\\ State Key Laboratory of Theoretical Physics(SKLTP) \\ Institute of
Theoretical Physics, Chinese Academy of Science, Beijing,100190, China}

\author{Yong Tang}\email{ytang@phys.cts.nthu.edu.tw}
\affiliation{Physics Division, National Center for Theoretical Sciences, Hsinchu}

\author{Yue-Liang Wu}\email{ylwu@itp.ac.cn}
\affiliation{Kavli Institute for Theoretical Physics China (KITPC)
\\ State Key Laboratory of Theoretical Physics(SKLTP) \\ Institute of
Theoretical Physics, Chinese Academy of Science, Beijing,100190, China}

\date{\today}

\begin{abstract}
The Higgs decay $H\to \gamma\gamma$ due to the virtual $W$-loop
effect is revisited in the unitary gauge by using the
symmetry-preserving and divergent-behavior-preserving loop
regularization method, which is realized in the four dimensional
space-time without changing original theory. Though the one-loop
amplitude of $H\to \gamma\gamma$ is finite as the Higgs boson in the
standard model has no direct interaction with the massless photons
at tree level, while it involves both tensor-type and scalar-type
divergent integrals which can in general destroy the gauge
invariance without imposing a proper regularization scheme to make
them well-defined. As the loop regularization scheme can ensure the
consistency conditions between the regularized tensor-type and
scalar-type divergent irreducible loop integrals to preserve gauge
invariance, we explicitly show the absence of decoupling in the
limit $M_W/M_H\to0$ and obtain a result agreed exactly with the
earlier one in the literature. We then clarify the discrepancy
between the earlier result and the recent one obtained by R.
Gastmans, S.L. Wu and T.T. Wu. The advantage of calculation in the
unitary gauge becomes manifest that the non-decoupling arises from
the longitudinal contribution of the $W$ gauge boson.
\end{abstract}
\pacs{}

\maketitle

\section{Introduction}
The Higgs decay into two photons, $H\to \gamma\gamma$, is one of the
golden channels to discover the Higgs particle at the LHC,
especially if the Higgs mass is smaller than 130 GeV
\cite{Djouadi:2005gi}. Thus, a consistent theoretical calculation
for this process is of great interest experimentally. In the
standard model, there are two major contributions: one from the top
loop and one from the W-boson loop. All of these contributions have
been calculated by different groups for several decades
\cite{Ellis,shifman1,Ioffe,Rizzo,Resnick:1973vg} and their results
were consistent with each other. In particular, the $W$-boson loop
contribution showed, according to the previous calculations, an
interesting feature that it does not meet the intuitive picture of
decoupling for the infinite Higgs mass, namely it does not describes
the phenomenon that the Higgs particle ceases to decay to any other
particles including photons when its mass tends to infinity $M_H\to
\infty$ (or equivalently in the limit of $M_W\to 0$ in the present
case for the standard model with spontaneous symmetry breaking).
This feature has been argued transparently using the equivalence
theorem\cite{shifman2}. Note that this non-decoupling behavior is
quite different from the usual one \cite{Appelquist} which is
related to another limit $M_H/M_W\to 0$ and can be understood
intuitively in the following way: in the standard model the Higgs
coupling to other fields is proportional to the masses of the
latter, thus compensating the loop mass suppression
\cite{Djouadi:2005gi,shifman1}.

Recently, R. Gastmans, S.L. Wu and T.T. Wu \cite{Gastmans:2011ks,Gastmans:2011wh} raised a question to the well-known earlier result presented in the literature\cite{Ellis,shifman1,Ioffe,Rizzo},
where they performed a new calculation for the Higgs decay $H\to \gamma\gamma$ with the $W$-boson loop contribution and obtained a different result with the earlier one. Their calculation was carried out in
the unitary gauge in which all of the propagating degrees of freedom
are physical, which is, as far as we know, the first calculation for
this process in the unitary gauge. Particularly, their result displayed the
decoupling behavior that the resulting amplitude tends to vanish in the limit
$M_W/M_H\to0$. As claimed in their article that no any specific regularization method was used and the main difference of their
result from the earlier one was traced back to the use of
dimensional regularization\cite{'tHooft} in the earlier papers\cite{Ellis,shifman1,Ioffe}. The crucial point for such a statement is that, instead of using the regularization scheme to make the divergent integrals well-defined, the authors in\cite{Gastmans:2011ks,Gastmans:2011wh} adopted the replacement
\begin{equation}\label{relation}
l_\mu l_\nu\to \frac{1}{4}g_{\mu\nu}l^2
\end{equation}
in their calculation to relate the tensor-type and scalar-type divergent integrals
\begin{eqnarray}
I_{0\mu\nu} &=& \int d^4 l \frac{l_\mu
l_\nu}{[l^2-M_W^2+2\alpha_1\alpha_2(k_1\cdot k_2)]^3}\ , \\
I^\prime_0 &=& \int d^4 l
\frac{l^2}{[l^2-M_W^2+2\alpha_1\alpha_2(k_1\cdot k_2)]^3}
\end{eqnarray}
which leads the non-decoupling term to vanish identically.

To clarify the discrepancy between the recent
calculation\cite{Gastmans:2011ks,Gastmans:2011wh} and the earlier
calculations\cite{Ellis,shifman1,Ioffe,Rizzo}, we shall revisit in
this note the Higgs decay into two photons, $H\to \gamma\gamma$,
with the virtual $W$-boson loop contribution in the unitary gauge.
For the convenience of comparison and also with the advantage of
judiciously routing the external momenta through the loop, we will
take the same loop momentum variable choices and also some useful
notations given in ref. \cite{Gastmans:2011ks,Gastmans:2011wh}. It
is unlike the consideration in
\cite{Gastmans:2011ks,Gastmans:2011wh}, we shall keep using a proper
regularization method as the calculation of the amplitude involves
the tensor-type and scalar-type divergent integrals.  Since the loop
regularization(LORE) method \cite{Wu:2003dd,wu1} has been realized
in four dimensional space-time to make the divergent integrals
well-defined and to preserve all symmetries of original theory as
well as to maintain the divergent behavior of original integrals, we
are going to carry out a complete calculation for the Higgs decay
$H\to \gamma\gamma$ in the unitary gauge by using the LORE method.
In fact, the consistency of the LORE method has been checked by performing many one-loop and
even some two-loop calculations in many typical physical systems. Such as, it has explicitly been proved at one loop level that
the LORE method can preserve non-Abelian gauge symmetry and recover the correct $\beta$ function of QCD
\cite{Cui:2008uv} and mountain supersymmetry \cite{Cui:2008bk},  and provide a
consistent calculation for the chiral anomaly\cite{Ma:2005md} and
the radiatively induced Lorentz and CPT-violating Chern-Simons term
in QED\cite{Ma:2006yc} as well as for the QED trace
anomaly\cite{cui:2011}, and it also allows us to derive the dynamically
generated spontaneous chiral symmetry breaking of the low energy QCD
for understanding the origin of dynamical quark masses and the mass
spectra of light scalar and pseudoscalar nonet mesons in a chiral
effective field theory\cite{DW}, and to carry out the quantum
gravitational contributions to gauge theories with asymptotic free
power-law running\cite{Tang:2008ah,Tang:2010cr,Tang2011}.
The consistency and advantage of the LORE method beyond one loop
order has further been demonstrated by merging with Bjorken-Drell's
analogy between Fynman diagrams and electric circuits and also by
explicitly applying to the two-loop regularization and
renormalization of $\phi^4$ theory\cite{Huang:2011xh}. As a
consequence, we explicitly show the absence of decoupling for
infinite Higgs mass and arrive at the result which agrees exactly
with the earlier one\cite{Ellis,shifman1,Ioffe,Rizzo}. It is also
manifest to see the advantage of using the unitary gauge in the
calculation that the non-decoupling term arises from the
longitudinal contribution of the virtual $W$ gauge boson, which is
consistent with the general discussions given recently
in\cite{shifman2,Marciano:2011gm,Jegerlehner:2011jm}. Note that as
the LORE method is exactly defined in four dimension space-time, so
it does not plague the question raised in
\cite{Gastmans:2011ks,Gastmans:2011wh} for the dimensional
regularization. Here we would like to emphasize that for divergent
integrals, either logarithmic or quadratic, one cannot in general
make any manipulation, including the replacement
Eq.(\ref{relation}), before imposing firstly a proper regularization
scheme to make them well-defined.

Furthermore, we will also see that the use of Dyson's prescription\cite{Dyson} in
\cite{Gastmans:2011ks,Gastmans:2011wh} to eliminate the
gauge-invariance-violating term is improper. In fact, when consistently dealing
with the divergent integrals by applying for the LORE method, there is
a finite term which has the opposite sign to the gauge-invariance-violating term, so that they can cancel each other exactly. Thus,
one does not need to impose Dyson's prescription at all.

The paper is organized as follows: In Sec. II, following the refs.
\cite{Gastmans:2011ks,Gastmans:2011wh}, we write the amplitudes for
the three relevant Feynman diagrams explicitly in the unitary gauge
and with the particular choice of loop momentum variables. Then we
apply the LORE method to calculate the decay amplitude. In Sec. III, we make
some comments on all the existing results and statements. Especially, we will clarify, from either the conceptual or the physical points of view
, the difference between our computed result and the one obtained in\cite{Gastmans:2011ks,Gastmans:2011wh}. We
also discuss the application of dimensional regularization to the
present problem. In the final Sec., we come to our conclusions.

\section{Calculation for $H\to \gamma\gamma$ Amplitudes with LORE Method}

\begin{figure}[ht]
\begin{center}
  \includegraphics[scale=0.4]{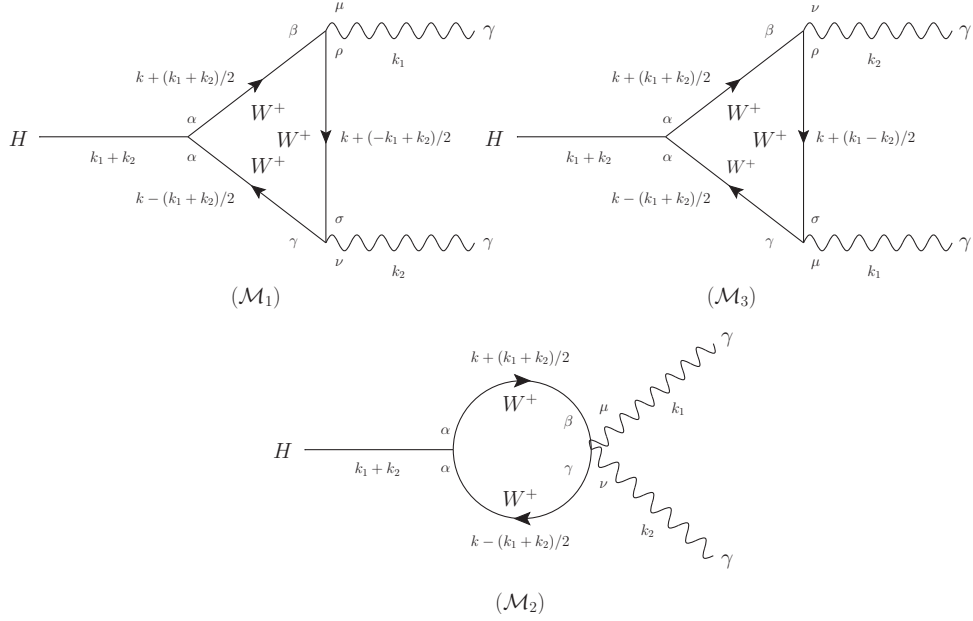}
  \caption{The one-loop diagrams with virtual W's in the unitary gauge that contribute to the amplitude for $H\to \gamma\gamma$}\label{H1}
\end{center}
\end{figure}

To make clarification for the recent result obtained in \cite{Gastmans:2011wh,Gastmans:2011ks}, we shall follow their calculation except for the
treatment on the tensor-type and scalar-type divergent integrals. Here we will properly
regularize the divergent integrals first by adopting the LORE method in which we have introduced the key concept of irreducible loop integrals(ILIs) and demonstrated the consistency conditions of gauge invariance among the regularized divergent ILIs\cite{Wu:2003dd,wu1}. We will work in the
unitary gauge in which any unphysical degrees of freedom do not
appear. Also, we particularly choose the loop momentum as in
Fig.(\ref{H1}), which enables the computation much simpler as shown in \cite{Gastmans:2011wh,Gastmans:2011ks}. By using
the Feynman rules listed in the appendix, it is straightforward to
write down the amplitudes for the three diagrams \cite{Gastmans:2011wh,Gastmans:2011ks}
\begin{eqnarray}
{\cal M}_1&=&\frac{-ie^2gM_W}{(2\pi)^4}\int d^4k\,[\,g_{\alpha}^{\beta}-(k+\dfrac{k_1+k_2}{2})_\alpha\,(k+\dfrac{k_1+k_2}{2})^\beta/M_W^2\,]\nonumber\\
&&\nonumber\\
&&\times[\,g^{\rho\sigma}-(k+\dfrac{-k_1+k_2}{2})^\rho\,(k+\dfrac{-k_1+k_2}{2})^\sigma/M_W^2\,]\nonumber\\
&&\nonumber\\
&&\times[\,g^{\alpha\gamma}-(k-\dfrac{k_1+k_2}{2})^\alpha\,(k-\dfrac{k_1+k_2}{2})^\gamma/M_W^2\,]  \\
&&\nonumber\\
&&\times[\,(k+\dfrac{3k_1+k_2}{2})_\rho\, g_{\beta\mu}+(k+\dfrac{-3k_1+k_2}{2})_\beta\, g_{\mu\rho}+(-2k-k_2)_\mu\, g_{\rho\beta}\,]\nonumber\\
&&\nonumber\\
&&\times\frac{(k-\dfrac{k_1+3k_2}{2})_\sigma\,g_{\gamma\nu}+(k+\dfrac{-k_1+3k_2}{2})_\gamma\,
g_{\nu\sigma}+(-2k+k_1)_\nu\, g_{\sigma\gamma}}
{\big[\big(k+\frac{k_1+k_2}{2}\big)^2-M_W^2+i\epsilon\big]\,\big[\big(k+\frac{-k_1+k_2}{2}\big)^2-M_W^2+i\epsilon\big]\,
\big[\big(k-\frac{k_1+k_2}{2}\big)^2-M_W^2+i\epsilon\big]}\,,\nonumber
\label{eq2.1}
\end{eqnarray}
\begin{eqnarray}
{\cal M}_2&=&\frac{ie^2gM_W}{(2\pi)^4}\int d^4k\,[\,g_{\alpha}^{\beta}-(k+\dfrac{k_1+k_2}{2})_\alpha\,(k+\dfrac{k_1+k_2}{2})^\beta/M_W^2\,]\nonumber\\
&&\nonumber\\
&&\times[\,g^{\alpha\gamma}-(k-\dfrac{k_1+k_2}{2})^\alpha\,(k-\dfrac{k_1+k_2}{2})^\gamma/M_W^2\,]  \\
&&\nonumber\\
&&\times\frac{2\,g_{\mu\nu}\,g_{\beta\gamma}-g_{\mu\beta}\,g_{\nu\gamma}-g_{\mu\gamma}\,g_{\nu\beta}}
{\big[\big(k+\frac{k_1+k_2}{2}\big)^2-M_W^2+i\epsilon\big]\,
\big[\big(k-\frac{k_1+k_2}{2}\big)^2-M_W^2+i\epsilon\big]}\,, \nonumber
\label{eq2.2}
\end{eqnarray}
and
\begin{eqnarray}
{\cal M}_3&=&\frac{-ie^2gM_W}{(2\pi)^4}\int d^4k\,[\,g_{\alpha}^{\beta}-(k+\dfrac{k_1+k_2}{2})_\alpha\,(k+\dfrac{k_1+k_2}{2})^\beta/M_W^2\,]\nonumber\\
&&\nonumber\\
&&\times[\,g^{\rho\sigma}-(k+\dfrac{k_1-k_2}{2})^\rho\,(k+\dfrac{k_1-k_2}{2})^\sigma/M_W^2\,]\nonumber\\
&&\nonumber\\
&&\times[\,g^{\alpha\gamma}-(k-\dfrac{k_1+k_2}{2})^\alpha\,(k-\dfrac{k_1+k_2}{2})^\gamma/M_W^2\,] \\
&&\nonumber\\
&&\times[\,(k+\dfrac{k_1+3k_2}{2})_\rho\, g_{\beta\nu}+(k+\dfrac{k_1-3k_2}{2})_\beta\, g_{\nu\rho}+(-2k-k_1)_\nu\, g_{\rho\beta}\,]\nonumber\\
&&\nonumber\\
&&\times\frac{(k-\dfrac{3k_1+k_2}{2})_\sigma\,g_{\gamma\mu}+(k+\dfrac{3k_1-k_2}{2})_\gamma\,
g_{\mu\sigma}+(-2k+k_2)_\mu\, g_{\sigma\gamma}}
{\big[\big(k+\frac{k_1+k_2}{2}\big)^2-M_W^2+i\epsilon\big]\,\big[\big(k+\frac{k_1-k_2}{2}\big)^2-M_W^2+i\epsilon\big]\,
\big[\big(k-\frac{k_1+k_2}{2}\big)^2-M_W^2+i\epsilon\big]}\,.\nonumber
\label{eq2.3}
\end{eqnarray}

With this choice of the loop momentum variables and applying Ward
identities, it was shown in \cite{Gastmans:2011wh,Gastmans:2011ks}
that there is a great cancelation in the integrand among these
three diagrams, without shifting the momentum variable or performing
the integrations. Especially without using the relation
\begin{equation}
l_\mu l_\nu\to \frac{1}{4}l^2 g_{\mu\nu} \nonumber \ ,
\end{equation}
the only remaining parts are those given by ${\cal M}_{1131}$, ${\cal M}_{1132}$,
${\cal M}_{123}$, ${\cal M}_{143}$, ${\cal M}_{15}$, ${\cal M}_{24}$
and their counterparts in the diagram ${\cal M}_3$, here we have used the same notations and definitions as the ones in \cite{Gastmans:2011wh,Gastmans:2011ks}). Below we relabel these
parts of amplitudes into ${\cal M}_{L}$ and ${\cal M}_T$, so that the total
amplitude is given by
\begin{equation}
{\cal M} = {\cal M}_1 +{\cal M}_2+{\cal M}_3 = {\cal M}_L +{\cal
M}_T = {\cal M}^{(1)}_L +{\cal M}^{(3)}_L +{\cal M}_T^{(1)} +{\cal
M}_T^{(3)}
\end{equation}
where
\begin{eqnarray}
{\cal M}^{(1)}_L &\equiv& {\cal M}_{1131} = \frac{-i e^2 g M_W}{(2\pi)^4}\frac{1}{M_W^2}  \nonumber\\ 
& &  \cdot\int
d^4k\frac{A^\prime}{[(k+\frac{k_1+k_2}{2})^2-M_W^2][(k+\frac{-k_1+k_2}{2})^2-M_W^2][(k-\frac{k_1+k_2}{2})^2-M_W^2]},\nonumber
\end{eqnarray}
with
\begin{eqnarray}
A^\prime &=& 4(k_1\cdot k_2)k_\mu k_\nu +2k^2 k_{2\mu}k_{1\nu} -4 k_\mu k_{1\nu}-4k_{2\mu}k_\nu (k\cdot k_1)\nonumber\\
&& g_{\mu\nu}[-2k^2(k_1\cdot k_2)+4(k\cdot k_1)(k\cdot k_2)],
\end{eqnarray}
and
\begin{eqnarray}\label{M_15}
{\cal M}_T^{(1)}&\equiv&{\cal M}_{15}+\frac{1}{2}{\cal M}_{24}+{\cal
M}_{123}+{\cal
M}_{143}+{\cal M}_{1132}  \\
&=& \frac{-ie^2 g M_W}{(2\pi)^4}\int d^4
k\frac{B}{[(k+\frac{k_1+k_2}{2})^2-M_W^2][(k+\frac{-k_1+k_2}{2})^2-M_W^2]
[(k-\frac{k_1+k_2}{2})^2-M_W^2]} \ ,\nonumber
\end{eqnarray}
with
\begin{eqnarray}
B &=& g_{\mu\nu} [-3k^2+3(k\cdot k_1)-3(k\cdot
k_2)-\frac{9}{2}(k_1\cdot k_2)+3M_W^2]\nonumber\\
&& +12k_\mu k_{\nu}+3k_{2\mu}k_{1\nu}-6k_\mu k_{1\nu}+6k_{2\mu}k_\nu \ .
\end{eqnarray}
The expressions for ${\cal M}^{(3)}_L$ and ${\cal M}^{(3)}_T$ are
same as the ones for ${\cal M}^{(1)}_L$ and ${\cal M}^{(1)}_T$
except for the interchange of indices $1\leftrightarrow 2$. The
reason for the subscripts `L' and `T' is that they have different
origins: ${\cal M}_L$ contains terms which are all related to at
least one longitudinal polarization in three internal $W$-boson
propagators, while ${\cal M}_T$ represents the rest terms. This can
be easily seen from the extra factor $1\over M^2_W$ before the
integration in ${\cal M}_L$ compared with ${\cal M}_T$. Note that
both parts involve the logarithmic divergence, so they need to be
regularized first before performing integration. Here we shall apply
the LORE method to treat the divergent integrals and carry out a
consistent calculation.

The calculation of ${\cal M}^{(1)}_{L}$ is straightforward by
applying for the LORE method. Combining the three factors in the denominator
with the help of Feynman parameters $\alpha_1$ and $\alpha_2$, we have
\begin{eqnarray}
D &=& k^2 +\alpha_1 k\cdot (k_1+k_2)-\alpha_2
k\cdot(k_1+k_2)+(1-\alpha_1-\alpha_2) k\cdot
(-k_1+k_2)-\frac{(k_1\cdot k_2)}{2}-M_W^2\nonumber\\
&=& [k-\frac{1}{2}(1-2\alpha_1)k_1+\frac{1}{2}(1-2\alpha_2)k_2]^2
+2\alpha_1\alpha_2 (k_1\cdot k_2)-M_W^2 \ .
\end{eqnarray}
With the shift of loop momentum variable
\begin{equation}\label{shift}
k=l+\frac{1}{2}(1-2\alpha_1)k_1-\frac{1}{2}(1-2\alpha_2)k_2 \ ,
\end{equation}
the amplitude ${\cal M}^{(1)}_{L}$ becomes
\begin{eqnarray}
{\cal M}^{(1)}_{L} = \frac{-i e^2 g M_W}{(2\pi)^4}\frac{1}{M_W^2}
\Gamma(3) \int^1_0 d\alpha_1 \int^{1-\alpha_1}_0 d\alpha_2 \int d^4
l \frac{A^\prime}{[l^2-M_W^2+2\alpha_1\alpha_2(k_1\cdot k_2)]^3} \ ,
\end{eqnarray}
where the factor $\Gamma(3)$ comes from the Feynman parametrization
and
\begin{eqnarray}
A^\prime &\equiv& 4( k_1\cdot k_2)[l_\mu l_\nu-\alpha_1\alpha_2 k_{2\mu}k_{1\nu}] +2 k_{2\mu} k_{1\nu}(l^2-2\alpha_1\alpha_2 k_1\cdot k_2)\nonumber\\
&& -4[(k_2\cdot l)l_\mu k_{1\nu}-\alpha_1\alpha_2 (k_1\cdot k_2)k_{2\mu} k_{1\nu}] -4[(k_1\cdot l)l_\nu k_{2\mu}-\alpha_1\alpha_2 (k_1\cdot k_2)k_{2\mu} k_{1\nu}]\nonumber\\
&& -2g_{\mu\nu} (k_1\cdot k_2)[l^2-2\alpha_1\alpha_2 k_1\cdot k_2]+4g_{\mu\nu}[l\cdot k_1 l\cdot k_2-\alpha_1\alpha_2(k_1\cdot k_2)^2]\nonumber\\
&=& 4(k_1\cdot k_2)l_\mu l_\nu +2k_{2\mu}k_{1\nu} l^2-4(k_2\cdot l) l_\mu k_{1\nu} \nonumber\\
&& -4(l\cdot k_1)k_{2\mu}l_\nu -2g_{\mu\nu}(k_1\cdot k_2)l^2 +4
g_{\mu\nu}l\cdot k_1 l\cdot k_2 \ ,
\end{eqnarray}
where we have ignored the terms with odd number of $l$ as their integrations vanish. Notice that the terms
that do not involve the integral variable $l$ cancel with each other
exactly. Then it is easy to rewrite the expression of ${\cal
M}_L^{(1)}$ into the sum of the irreducible loop integrals (ILI's) introduced in the LORE method
\begin{eqnarray}
{\cal M}^{(1)}_{L} &=& \frac{-i e^2 g M_W}{(2\pi)^4}\frac{1}{M_W^2}
\Gamma(3) \int^1_0 d\alpha_1 \int^{1-\alpha_1}_0
d\alpha_2\nonumber\\
&& \{4[(k_1\cdot k_2)g^{\rho}_\mu g^\sigma_\nu -k_2^\sigma
k_{1\nu}g_\mu^\rho -k_1^\rho k_{2\mu}g^\sigma_\nu+ k_1^\rho
k_2^\sigma g_{\mu\nu}] I_{0\mu\nu}
+2[k_{2\mu}k_{1\nu}-g_{\mu\nu}(k_1\cdot k_2)]I_0\nonumber\\
&& +2[k_{2\mu}k_{1\nu}-g_{\mu\nu}(k_1\cdot
k_2)][M_W^2-2\alpha_1\alpha_2(k_1\cdot k_2)] I_{-2}\} \ ,
\end{eqnarray}
where we have defined the ILI's $I_0$, $I_{0\mu\nu}$ and $I_{-2}$ as
\begin{eqnarray}
I_{0} &=& \int d^4 l \frac{1}{[l^2-M_W^2+2\alpha_1\alpha_2(k_1\cdot k_2)]^2}\label{scalar0} \ ,\\
I_{0\mu\nu} &=& \int d^4 l \frac{l_\mu l_\nu}
{[l^2-M_W^2+2\alpha_1\alpha_2(k_1\cdot k_2)]^3}\label{tensor} \ ,\\
I_{-2} &=& \int d^4 l \frac{1}{[l^2-M_W^2+2\alpha_1\alpha_2(k_1\cdot
k_2)]^3}\nonumber\\
&=& -\frac{i\pi^2}{2}\frac{1}{[M_W^2-2\alpha_1\alpha_2(k_1\cdot
k_2)]} \ .
\end{eqnarray}
By applying for the LORE method and its resulting
consistency condition of gauge invariance \cite{wu1, Wu:2003dd}:
\begin{equation}\label{consistent}
I^R_{0\mu\nu} = \frac{1}{4} I^R_0 g_{\mu\nu} \ ,
\end{equation}
where $I^R_{0}$ is calculated as
\begin{equation}
I^R_0 = i\pi^2[\ln\frac{M_c^2}{M_W^2-2\alpha_1\alpha_2(k_1\cdot
k_2)}-\gamma_\omega+y_0(\frac{M_W^2-2\alpha_1\alpha_2(k_1\cdot
k_2)}{M_c^2})] \ ,
\end{equation}
and the function $y_0(x)\to0$ with $x\to0$ rapidly enough, we find
that the divergent integrals cancel each other and arrive at the
finite result
\begin{eqnarray}\label{M_L}
{\cal M}^{(1)}_L &=& \frac{-i e^2 g M_W}{(2\pi)^4}\frac{\Gamma(3)}{M_W^2}
 \int^1_0 d\alpha_1 \int^{1-\alpha_1}_0 d\alpha_2
2[k_{2\mu}k_{1\nu}-g_{\mu\nu}(k_1\cdot
k_2)][M_W^2-2\alpha_1\alpha_2(k_1\cdot k_2)] I_{-2}\nonumber\\
&=& \frac{-ie^2 g M_W}{(2\pi)^4}\frac{\Gamma(3)}{M_W^2} \frac{1}{2}
2[k_{2\mu}k_{1\nu}-g_{\mu\nu}(k_1\cdot
k_2)](\frac{-i\pi^2}{2})\nonumber\\
&=& -\frac{e^2 g}{8\pi^2
M_W}\frac{1}{2}[k_{2\mu}k_{1\nu}-g_{\mu\nu}(k_1\cdot k_2)] \ ,
\end{eqnarray}
where in the second line we have explicitly integrated out the finite integral $I_{-2}$ and the factor $1\over 2$ comes from the Feynman parameter
integrations.

We shall proceed to calculate the remaining term ${\cal M}_T^{(1)}$
in Eq.(\ref{M_15}). By the same Feynman parametrization and the same
shift of loop momentum variable as in Eq.(\ref{shift}), the
integration in Eq.(\ref{M_15}) becomes
\begin{eqnarray}
{\cal M}_T^{(1)}&=& \frac{-ie^2 g M_W}{(2\pi)^4} \Gamma(3)\int^1_0
d\alpha_1 \int^{1-\alpha_1}_0 d\alpha_2 \int d^4 l
\{\frac{-3g_{\mu\nu}l^2+12 l_\mu
l_\nu}{[l^2-M_W^2+2\alpha_1\alpha_2(k_1\cdot
k_2)]^3}\nonumber\\
&& +\frac{g_{\mu\nu}[(k_1\cdot
k_2)(-6+6\alpha_1\alpha_2)+3M_W^2]+3(2-4\alpha_1\alpha_2)k_{2\mu}
k_{1\nu}}{[l^2-M_W^2+2\alpha_1\alpha_2(k_1\cdot k_2)]^3}\}\nonumber\\
&=& \frac{-ie^2 g M_W}{(2\pi)^4} \Gamma(3)\int^1_0 d\alpha_1
\int^{1-\alpha_1}_0 d\alpha_2 \{(-3)(g_{\mu\nu}I_0-4
I_{0\mu\nu})\nonumber\\
&& +6[k_{2\mu}k_{1\nu}-g_{\mu\nu}(k_1\cdot
k_2)](1-2\alpha_1\alpha_2)I_{-2}\} \ ,
\end{eqnarray}
for which we shall use again the consistency condition Eq.(\ref{consistent}) for the regularized ILIs and find
that the gauge-invariance-violating term vanishes.  After carrying out the integration on $I_{-2}$, we obtain the finite result
\begin{eqnarray}\label{M_T}
{\cal M}_T^{(1)} &=& \frac{-e^2 g M_W}{16\pi^2} \int^1_0 d\alpha_1
\int^{1-\alpha_1}_0 d\alpha_2
\frac{6(1-2\alpha_1\alpha_2)[k_{2\mu}k_{1\nu}-g_{\mu\nu}(k_1\cdot
k_2)]}{M_W^2-2\alpha_1\alpha_2(k_1\cdot k_2)}
\end{eqnarray}

There are also two similar contributions ${\cal M}^{(3)}_L$ and ${\cal
M}_T^{(3)}$ from ${\cal M}_{3}$ including the other half of ${\cal
M}_2$, which are given by the same results as Eqs.(\ref{M_L}) and (\ref{M_T})
because they are just involving the interchange of indices
$1\leftrightarrow 2$. Thus, by doubling the sum of Eqs.(\ref{M_L})
and (\ref{M_T}), we arrive at the final result for the amplitude of the Higgs decay into two photons via $W$-loop
\begin{eqnarray}\label{final}
{\cal M} &=& {\cal M}_L^{(1)}+{\cal M}_T^{(1)}+{\cal M}^{(3)}_L+{\cal M}^{(3)}_T\nonumber\\
&=& -\frac{e^2 g}{16\pi^2 M_W}[k_{2\mu}k_{1\nu}-g_{\mu\nu}(k_1\cdot
k_2)][2+3\tau^{-1}+(2\tau^{-1}-\tau^{-2})f(\tau)] \ ,
\end{eqnarray}
where we have used the definitions
\begin{eqnarray}
\tau &\equiv& \frac{M_H^2}{4 M_W^2} \ ,\\
 f(\tau) &=& \left\{ \begin{array}{cc} \arcsin^2(\sqrt{\tau}) & for ~
 \tau\leq 1 \\
 -\frac{1}{4}\big[\ln\frac{1+\sqrt{1-\tau^{-1}}}{1-\sqrt{1-\tau^{-1}}}-i\pi\big]^2
 & for ~ \tau > 1 \end{array} \right.
\end{eqnarray}
It is manifest that our above result agrees with the earlier one\cite{shifman1,Ellis,Ioffe,Rizzo}.

\section{Some Comments and Remarks}

    With the above explicit calculation in the unitary gauge for the Higgs decay into two photons $H\to \gamma\gamma$ via the virtual $W$-loop, we are now in the position to make some comments and remarks for the result obtained in \cite{Gastmans:2011wh,Gastmans:2011ks}.

\subsection{The Conceptual Reason for the Discrepancy}

As mentioned in the introduction that the authors in
\cite{Gastmans:2011wh,Gastmans:2011ks} obtained a different result
from the earlier one, the reason for this discrepancy
is their use of the replacement
\begin{equation}
l_\mu l_\nu \to \frac{1}{4}l^2 g_{\mu\nu} \nonumber
\end{equation}
for the divergent integrals. Such a replacement is in general not valid for the divergent integrals as it may destroy gauge invariance through spoiling the consistency conditions between the regularized tensor-type and scalar-type ILIs. In fact, it was such a replacement used in\cite{Gastmans:2011wh,Gastmans:2011ks} to deal with the divergent tensor-type ILI $I_{0\mu\nu}$, that led to a result different from our above result and also the earlier one. Explicitly, when imposing such a replacement to the divergent tensor-type ILI $I_{0\mu\nu}$, one will yield the following relation
\begin{eqnarray}\label{wrong}
I_{0\mu\nu} = \frac{1}{4}g_{\mu\nu} I^\prime_0 \ ,
\end{eqnarray}
where the divergent integral $I^\prime_0$ is given by
\begin{eqnarray}
I^\prime_0 &\equiv& \int d^4 l
\frac{l^2}{[l^2-M_W^2+2\alpha_1\alpha_2 (k_1\cdot k_2)]^3}
\end{eqnarray}
Obviously, the relation eq.(\ref{wrong}) distinguishes from the consistency condition Eq.(\ref{consistent}) between the regularized tensor-type ILI $I^R_{0\mu\nu}$ and the scalar-type one $I^R_0$
\begin{equation}\label{scalar2}
I^R_{0\mu\nu} = \frac{1}{4} g_{\mu\nu}I^R_0 \nonumber
\end{equation}
This is because $I^R_{0\mu\nu}$ differs from $I^{\prime R}_0$ by a finite term. In fact,
the integral $I^{\prime}_0$ is not an ILI which was introduced as a key conceptual point in the LORE method. Explicitly, $I^{\prime}_0$ can be reexpressed into the sum of ILI's
\begin{eqnarray}
I^{\prime}_0 = I_0 +I_{-2},\quad \mbox{and} \quad  I^{\prime R}_0 = I_0^R +I_{-2}
\end{eqnarray}
which becomes manifest that it is the extra finite integral $I_{-2}$ that
makes the discrepancy between our above result and the one obtained in \cite{Gastmans:2011wh,Gastmans:2011ks}. Here we would like to emphasize that by taking into account of the divergence nature of Eqs. (\ref{tensor}) and (\ref{consistent}),
we cannot say anything before regularizing them
properly. Only after making them well defined through regularization,
we can then perform the ordinary manipulations in 4-dimensions, including the
replacement Eq.(\ref{relation}), which ultimately
leads to the consistency condition Eq.(\ref{consistent})

The lack of $I_{-2}$ in the derivation \cite{Gastmans:2011wh,Gastmans:2011ks} caused two consequences: firstly, the obtained result displayed the decoupling behavior. Secondly, it led to the appearance of
the gauge-invariance-violating term which was argued to be eliminated by the use of Dyson's
prescription. We are going to discuss these two points further below.

\subsection{The Non-Decoupling Contribution}

In the limit $M_W/M_H\to 0$, the decay amplitude given in Eq.(\ref{final})
becomes:
\begin{equation}\label{non-decouple}
{\cal M}_{non-decouple} = -\frac{e^2 g}{8\pi^2
M_W}[k_{2\mu}k_{1\nu}-g_{\mu\nu}(k_1\cdot k_2)] \ ,
\end{equation}
which implies the absence of the decoupling in the limit of infinite
Higgs mass. Thus our present calculation in the unitary gauge
confirms the earlier one\cite{shifman1, Ellis, Ioffe, Rizzo} and is
consistent with the conclusion by using the argument of the
equivalence theorem\cite{shifman1,shifman2,Korner:1995xd}.

Let us make a further connection for our present result obtained in the unitary gauge with the general argument based on
the equivalence theorem. From the above explicit formulation, it is seen that the
non-decoupling term Eq.(\ref{non-decouple}) can be traced back to
the part ${\cal M}_L={\cal M}_L^{(1)}+{\cal M}_L^{(3)}$, which
involves only the longitudinal parts of the propagator
$\frac{1}{p^2-M_W^2}\frac{p^\alpha p^\beta}{M^2_W}$. This indicates that it is the longitudinal polarization that prevents the Higgs
boson decoupling from the photons, which is one of the advantages calculating in the
unitary gauge in which all the propagating degrees of freedom are
physical, so that we can easily identify the origin of the non-decoupling contributions.

The relationship between the non-decoupling contribution and the
longitudinal polarization of the W-boson can become more apparent with
the help of the equivalence theorem \cite{shifman2,shifman1}. Recall
that in the standard model the $W$-boson mass is related to the vacuum
expectation value (VEV) $v$ via
\begin{equation}
M_W = \frac{g v}{2}
\end{equation}
with $g$ the SU(2)$_L$ gauge coupling. When fixing $v$ and
$M_H$, the limit $M_W/M_H\to 0$ corresponds to a vanishing SU(2)$_L$
coupling $g\to0$, thus the transverse polarizations of $W$-boson
fully decouple from the Higgs boson. The only relevant degrees of
freedom are the longitudinal polarization parts. According to the
equivalence theorem\cite{Vanishtein,Cornwall,Chanowitz}, the Higgs
coupling to the longitudinal polarized $W$-bosons is equivalent to
the coupling to the massless Goldstone bosons, which are known to be `eaten'
by the $W$-boson via the Higgs mechanism. Two of the three Goldstone
bosons $\phi^+$ and $\phi^-$ are charged and mediate the $H\to
\gamma\gamma$ decay due to the effective interaction with the Higgs
boson
\begin{equation}
{\cal L}_{HWW}\to -\frac{M_H^2}{v}H\phi^+\phi^-
\end{equation}
which leads to non-decoupling contributions.

\subsection{Elimination of Gauge-Invariance-Violating Term}

As shown in\cite{Gastmans:2011ks,Gastmans:2011wh} that the amplitude ${\cal
M}_T^{(1)}$ got the following result ( see Eq.(3.50) in refs.\cite{Gastmans:2011ks,Gastmans:2011wh} ) after using the replacement eq.(\ref{relation})
\begin{eqnarray}
{\cal M}_{T}^{(1)} &=& \frac{-e^2 g M_W}{16\pi^2} \int^1_0 d\alpha_1
\int^{1-\alpha_1}_0 d\alpha_2 \frac{g_{\mu\nu}[(k_1\cdot
k_2)(-6+6\alpha_1\alpha_2)+3M_W^2]+3(2-4\alpha_1\alpha_2)k_{2\mu}k_{1\nu}}
{M_W^2-2\alpha_1\alpha_2(k_1\cdot k_2)}\nonumber\\
&=& \frac{-e^2 g M_W}{16\pi^2} \int^1_0 d\alpha_1
\int^{1-\alpha_1}_0 d\alpha_2
\{\frac{6(1-2\alpha_1\alpha_2)[k_{2\mu}k_{1\nu}-g_{\mu\nu}(k_1\cdot
k_2)]}{M^2-2\alpha_1\alpha_2(k_1\cdot k_2) }+3g_{\mu\nu}\}\ ,
\end{eqnarray} 
which is compared to our result Eq.(\ref{M_T}) with an extra term $3g_{\mu\nu}$ in the parenthesis. Note that this extra term
will break the gauge invariance in the theory as it is only proportional to $g_{\mu\nu}$ in the final result. In order to eliminate
such an unwanted term, it was suggested in \cite{Gastmans:2011wh,Gastmans:2011ks} to subtract it
directly by the argument of Dyson's prescription \cite{Dyson}. It is actually improper to use here the Dyson's prescription which
was originally motivated to relate the concept of
on-shell renormalization by performing the subtraction of the amplitude
at zero external momentum. However, for the processes like
$H\to\gamma\gamma$, the one-loop contribution is finite, so it does
not require renormalization (or subtraction) at all. The direct
calculation of the Feynman diagrams will naturally lead to a
consistent result when preserving the symmetries of the
underlying theory, like gauge invariance, which is proportional to the
special external momentum combination
$[k_{2\mu}k_{1\nu}-g_{\mu\nu}(k_1\cdot k_2)]$ in the present case.
Obviously, the appearance of a term proportional to
$g_{\mu\nu}$ only indicates the inconsistency of the
calculation. As it can be seen in our above calculation which does not involve this
`pathology'. An extra finite term arises as the remanent of the
cancelation between the divergent integrals
\begin{eqnarray}
g_{\mu\nu}I^{\prime R}_0-4I_{0\mu\nu}^R =
g_{\mu\nu}[M_W^2-2\alpha_1\alpha_2(k_1\cdot k_2)]I_{-2} =
(-\frac{i\pi^2}{2})g_{\mu\nu} \ ,
\end{eqnarray}
which consistently eliminates the unpleasant term mentioned above. Such a cancelation
again reflects the necessity to regularize the divergences properly, even for the logarithmic
divergences.

\subsection{Result with Dimensional Regularization}

From the above analysis, it becomes manifest for the reasons causing the discrepancy between our present result and the one obtained in \cite{Gastmans:2011wh,Gastmans:2011ks} in the same unitary gauge but with different treatment for the divergent integrals. As an independent check, we also carry out a calculation by using the dimensional regularization and arrive at the same result. In fact, the crucial point is the use of the consistency condition
Eq.(\ref{consistent}) and the relation
$I^{R\prime}_0=I^R_0+I_{-2}$ in the calculation. The
validity of these two formula has already been demonstrated for both the LORE method and the dimensional regularization as they only involve the logarithmic divergence \cite{Wu:2003dd,wu1}, it is then natural
to arrive at the same result Eq.(\ref{final}) in the dimensional regularization. While the difference between the LORE method and the dimensional regularization can arise in the treatment for the quadratic divergences\cite{Wu:2003dd,wu1,Huang:2011xh,DW,Tang:2008ah,Tang:2010cr,Tang2011}.

\section{Conclusions}

In this note, we have performed a calculation for the amplitude of the
process $H\to \gamma\gamma$ through a $W$-boson loop in the unitary
gauge by using the LORE method to treat consistently the tensor-type and scalar-type divergent integrals. Our present result given in Eq.(\ref{final}) has been found to agree exactly with the earlier one computed by several groups
\cite{shifman1, Ellis, Ioffe, Rizzo} in different approaches. We have clarified from an explicit calculation the discrepancy between our present result and the one obtained recently in\cite{Gastmans:2011wh,Gastmans:2011ks} in the same unitary gauge, and confirmed the well-known earlier result in the literature. In particular, we have shown the absence of the decoupling behavior in the limit of infinite Higgs mass $M_H\to \infty$ which is different from claim in\cite{Gastmans:2011wh,Gastmans:2011ks}. Here we would like to emphasize that it is necessity to make the divergent integrals, either logarithmic or quadratic, be well-defined through proper regularization schemes before carrying out any manipulations, including the replacement in Eq.(\ref{relation}), otherwise it may lead to an inconsistent result and conclusion. The calculation for the finite amplitude of the
process $H\to \gamma\gamma$ through a $W$-boson loop in the unitary
gauge has provided a good example to demonstrate that it is crucial for all the regularization schemes to satisfy the consistency conditions between the regularized tensor-type and scalar type divergent ILIs, which has been proved as a key concept in the LORE method.

\vspace{1 cm}

\centerline{{\bf Acknowledgement}}
\vspace{20 pt}
The author (Y. Tang) would like to thank C.-Q. Geng and L.-F. Li for helpful discussions.
This work is partly supported by the National Science
Foundation of China (NSFC) under Grant \#No. 10821504, 10975170 and
the key project of the Chinese Academy of Sciences.

\newpage

\appendix
\section{The Relevant Feynman Rules in the Unitary Gauge for the Decay $H\to\gamma\gamma$ Through one W-Boson Loop}
\begin{figure}[ht]
\begin{center}
  \includegraphics[scale=0.6]{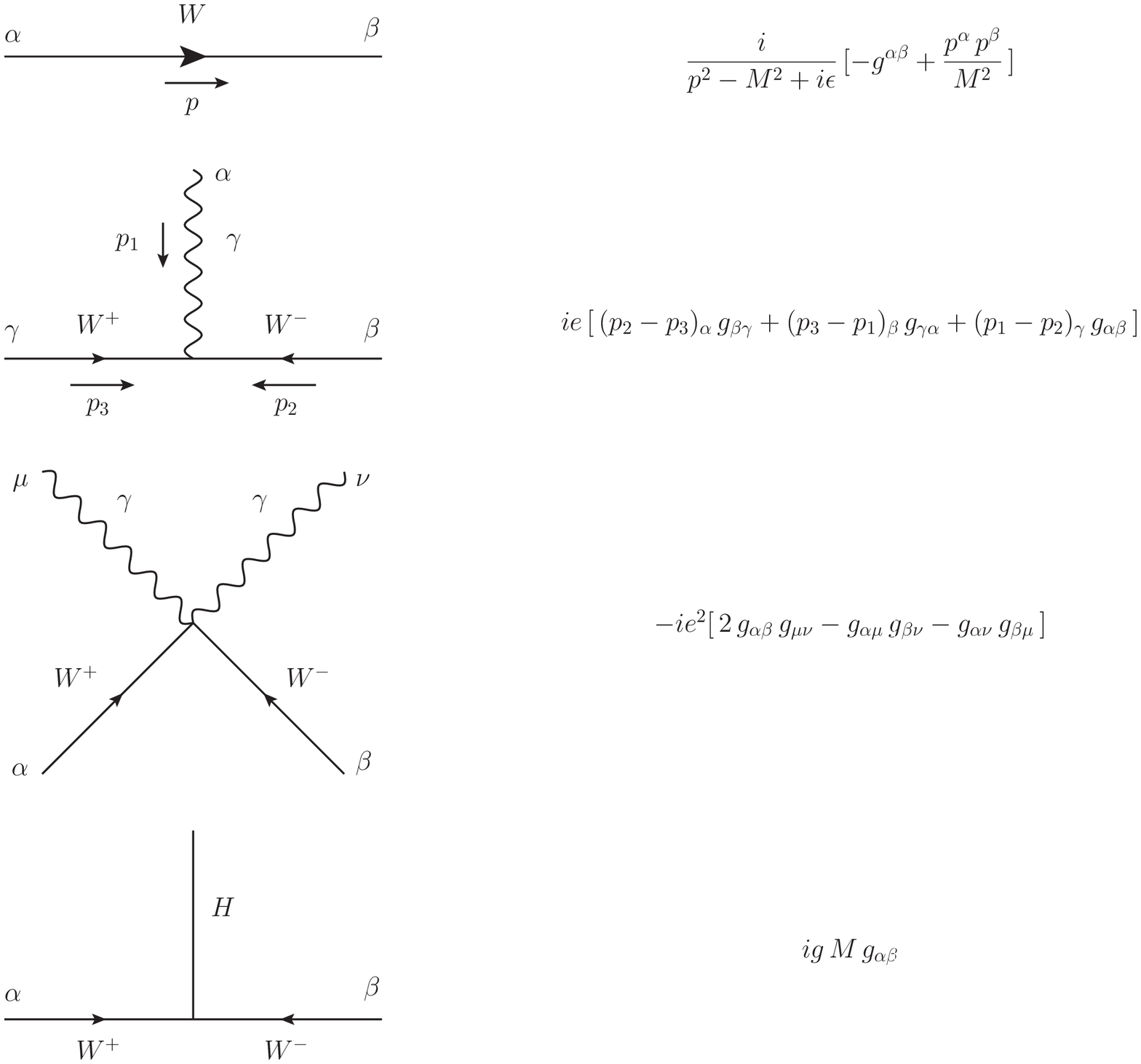}
  \caption{}\label{feyrule}
\end{center}
\end{figure}

\end{document}